\documentclass[12pt]{extarticle}
\usepackage[T1]{fontenc}
\usepackage[latin9]{inputenc}
\usepackage{geometry}
\usepackage{float}
\usepackage{amsmath}
\usepackage{graphicx}
\usepackage{setspace}
\makeatletter

\providecommand{\tabularnewline}{\\}

\usepackage[bottom]{footmisc}% places footnotes at page bottom

\usepackage[style=authoryear,natbib=true,firstinits=true,backend=bibtex]{biblatex}
\renewbibmacro{in:}{}

\usepackage{txfonts}
\usepackage{upgreek}

\usepackage{newunicodechar} 

\usepackage{bbm}

\newcommand{\prox}{\mbox{prox}}

\newcommand{\iterc}{{\kappa}}
\usepackage{algorithm}
\usepackage{algorithmic}

\makeatother

\begin{document}

\title{Generalized Linear Model for Gamma Distributed Variables via Elastic
Net Regularization }

\author{Xin Chen, Aleksandr Y. Aravkin, and R. Douglas Martin}
\maketitle
\begin{abstract}
The Generalized Linear Model (GLM) for the Gamma distribution (glmGamma)
is widely used in modeling continuous, non-negative and positive-skewed
data, such as insurance claims and survival data. However, model selection
for GLM depends on AIC/BIC criteria, which is computationally impractical
for even a moderate number of variables. In this paper, we develop
variable selection for glmGamma using elastic net regularization (glmGammaNet),
for which we provide an algorithm and implementation. The glmGammaNet
model is more challening than other more common GLMs as the likelihood
function has no global quadratic upper bound, and we develop an efficient
accelerated proximal gradient algorithm using a local model. We report
simulation study results and discuss the choice of regularization
parameter. The method is implemented in the R package glmGammaNet.
\end{abstract}

\section{Introduction}

Generalized Linear Models (GLMs) \citep{mccullagh1989generalized}
are used for inference when outcomes are binary, multinomial, count,
or non-negative. Regularization plays a key role for many GLM formulations;
in particular the $\ell_1$ norm \citep{tibshirani1996regression}
and elastic net \citep{zou2005regularization}, a linear combination
of the $\ell_1$ and quadratic loss, are frequently used to select
the most important predictors and predictor groups from a large set
of candidate variables. 

We focus on GLM models with Gamma-distributed response variables (i.e.
the responses are non-negative). This work is motivated by a recent
effort to estimate the standard errors of nonparametric sample estimators
for risk and performance measures \citep{chen2017standarderror}.
Estimation and cross-validation requires many evaluations of glmGammaNet,
so the approach must be parallelizable. There is currently no R package
that implements a parallelizable GLM for Gamma, so the current work
fills this gap. Table \ref{tab:Comparison-of-R-GLM} is a summary
of existing R packages for GLM, to the authors' best knowledge. In
particular, we provide an efficient, parallelizable package that can
fit a GLM model with EN regularization for the Gamma family. 

\begin{table}[th]
\begin{centering}
\begin{tabular}{|c|c|c|c|c|}
\hline 
{\footnotesize{}Package} & {\footnotesize{}Function} & {\footnotesize{}Support Gamma Dist} & {\footnotesize{}Model Selection} & {\footnotesize{}Multicore Parallel}\tabularnewline
\hline 
\hline 
{\footnotesize{}glmnet} & {\footnotesize{}glmnet()} & {\footnotesize{}No} & {\footnotesize{}EN Regularization} & {\footnotesize{}Yes}\tabularnewline
\hline 
{\footnotesize{}h2o} & {\footnotesize{}h2o.glm()} & {\footnotesize{}No} & {\footnotesize{}EN Regularization} & {\footnotesize{}No}\tabularnewline
\hline 
{\footnotesize{}stats} & {\footnotesize{}glm()} & {\footnotesize{}Yes} & {\footnotesize{}AIC/BIC} & {\footnotesize{}Yes}\tabularnewline
\hline 
{\footnotesize{}bestglm} & {\footnotesize{}bestglm()} & {\footnotesize{}No} & {\footnotesize{}Subset AIC/BIC} & {\footnotesize{}Yes}\tabularnewline
\hline 
\textbf{\footnotesize{}glmGammaNet} & \textbf{\footnotesize{}glmGammaNet()} & \textbf{\footnotesize{}Yes} & \textbf{\footnotesize{}EN Regularization} & \textbf{\footnotesize{}Yes}\tabularnewline
\hline 
\end{tabular}
\par\end{centering}
\caption{Comparison of R implementations for GLM\label{tab:Comparison-of-R-GLM}}
\end{table}

The optimization problem for the Gamma family is more challenging
than that for linear or logistic regression. The objective function
required to perform the inference does not have a global quadratic
upper bound. Such bounds are very useful for designing simple and
efficient first-order methods for penalized log-likelihood estimation.
Without the bound, a line search is needed in theory to ensure descent.
Instead, we estimate a quadratic bound locally using the functional
form of the Gamma to get a fast and robust method for the problem.
We implement a safeguard line search, but it is never activated. 

The paper proceeds as follows. In Section ~\ref{sec:rel_work}, we
give a brief survey of GLM use cases and algorithms. We also discuss
the role of regularization and its impact on choice of algorithm.
In Section ~\ref{sec:algo}, we formulate the Gamma inference problem,
dicuss first-order methods for elastic-net (EN) regularization, and
detail the algorithm we implemented. Section~\ref{sec:sim} presents
simulation results.  We end with a discussion in Section~\ref{sec:disc}.

\subsection{\label{sec:rel_work} Related work}

First introduced by \citet*{glm1972}, GLM has been used for variety
of applications, including Binary logistic regression, Multinomial
logistic regression, ordinal logistic regression and Poisson regression,
see for example \citealp{mccullagh1989generalized} and \citealp*{dobson2008introduction}.
Gamma GLMs are used to model right-skewed non-negative data, such
as insurance claims \citep*{JongPietde2008Glmf}, Semiconductor Wafer
sensitivity \citep{myers1997tutorial}, clotting times of blood \citep*{mccullagh1989generalized}
and Survival Function of Diabetic Nephropathy Patients \citep{grover2013application}.
In R, GLMs are often fit using the glm() function in stats package.

Model selection is essential for GLMs. While classic approaches use
AIC/BIC criteria (see e.g. \citealp{bozdogan1987aicmodel,burnham2003aicmodel,burnham2004aicbicmultimodel}),
sparsity-based regularization is very useful \citep*{tibshirani1996regression,zou2005regularization}\textbf{. }

\section{\label{sec:algo} First-Order Methods for Regularized GLM}

The GLM inference problem is formulated as follows. Suppose we wish to predict an output $b$ of a certain system on an input $a\in\mathbb{R}^n$. Let us also make the following two assumptions: $(i)$ the relationship between the input $a$ and the output $b$ is fairly simple and $(ii)$ we have available examples $a_i \in \mathbb{R}^n$ together with inexactly observed responses $b_i$ for  $i=1,\ldots,m$. 
The tuples $\{(b_i,a_i)\}_{i=1}^m$ comprise the training data.  The responses ${b}_i$ can have special restrictions; 
for example they may be counts, indicate class membership, or be non-negative, such as concentration of sugar in the blood.  To build the GLM, suppose the distribution of ${b}_i$  is parametrized by $(\mu_i, \sigma^2)$: 
\[ L(b_i|\mu_i, \sigma^2) = g_1(b_i, \sigma^2) \exp\left(\frac{b_i\mu_i - g_2(\mu_i)}{g_3(\sigma^2)}\right). \] 
To obtain the GLM objective, set $\mu_i := \langle a_i, x\rangle$, and take the negative log-likelihood 
(igore $g_1$ and $g_3$ as they do not depend on $x$): 
\begin{equation}
\label{eq:GLM}
\min_{x} \sum_{i=1}^m g_2(\langle a_i, x\rangle) - b_i\langle a_i, x\rangle
\end{equation}
  Common examples are shown in Table 2.

\begin{center}
\begin{table}[th]
\begin{centering}
\begin{tabular}{|c|c|c|}
\hline 
Model & Restriction on $b_i$ & $g_2(z)$\tabularnewline
\hline 
\hline 
Regression & None & $\frac{1}{2}\|z\|^2$\tabularnewline
\hline 
Classification & $b_i \in \{0,1\}$ & $\log(1+\exp(z))$\tabularnewline
\hline 
Counts & $b_i \in \mathbb{Z}_+$ & $\exp(z)$\tabularnewline
\hline 
Non-negative & $b_i \geq 0$ & $-\ln(z)$\tabularnewline
\hline 
\end{tabular}
\par\end{centering}
\caption{Common Generalized Linear Models}
\end{table}
\par\end{center}

We are interested in an extension of~\eqref{eq:GLM} that includes
nonsmooth regularizaiton terms $R(x)$ (inluding 1-norm, elastic net,
or constraints): 

\begin{equation}
\label{eq:GLM-reg}
\min_{x} \sum_{i=1}^m g_2(\langle a_i, x\rangle) - b_i\langle a_i, x\rangle + R(x)
\end{equation}

A simple strategy for optimizing models of form~\eqref{eq:GLM-reg}
is to develop simple upper bounds and minimize them. If $g$ is smooth,
its gradient is said to satisfy the Lipschitz property with constant
$L$ if

\begin{equation}
\label{eq:Lipschitz}
\|\nabla g(x) - \nabla g(y)\| \leq L \|x-y\| \quad \forall x,y.
\end{equation}

When $g$ is twice continuously differentiable, any bound on the operator
norm of $\nabla^2 g$ is a Lipschitz constant for $\nabla g$. For
example, if $g(x) = \frac{1}{2}\|Ax -b\|^2$, the Lipschitz constant
for $\nabla g(x) = A^T(Ax-b)$ is the largest eigenvalue of $A^TA$.
Any Lipschitz constant for $\nabla g$ gives a simple tight global
upper bound for $g$:

\[
g(x) \leq g(x_0) + (x-x_0)^T \nabla g(x_0) + \frac{L}{2}\|x-x_0\|^2.
\]

In the context of model~\eqref{eq:GLM-reg}, suppose that $\nabla g_2$
has Lipschitz constant $l$. Then if we define 

\[
g(x) = \sum_{i=1}^m g_2(\langle a_i, x\rangle) - b_i\langle a_i, x\rangle,
\]

the reader can immediately see that the Lipschitz constant of $\nabla g$
is bounded above by $l\|A\|^2$. A simple iterative strategy is to
minimize the upper bound for $g$ at each iteration, without modifying
$R(x)$, which may be non-smooth (1-norm) or infinite valued (box-constraint).
Given iterate $x^k$, the next iterate $x^+$ is found as follows:

\begin{equation}
\label{eq:prox_def}
\begin{aligned}
x^+ &= \arg\min_{x} (x-x^k)^T \nabla g(x^k) + \frac{L}{2}\|x-x^k\|^2 + R(x) \\
&= \arg\min_x \frac{1}{2L} \|x-(x^k-\frac{1}{L}\nabla g(x^k)\|^2 + R(x) \\
&:= \mbox{prox}_{\frac{1}{L}R} (x^k-\frac{1}{L}\nabla g(x^k)).
\end{aligned}
\end{equation}

The proximity operator $\mbox{prox}_{\frac{1}{L}R}(z)$ defined in~\eqref{eq:prox_def}
should be thought of as a simple subroutine. It is the minimizer of
the problem 

\[
\min_{x} \frac{1}{2L} \|x-z\|^2 + R(x),
\]

and is available in closed form for a wide variety of regularizers
$R(x)$, including 1-norm , the elastic net, and simple constraints
\citep{combettes2011proximal}. 

\begin{algorithm} \caption{FISTA for Regularized GLM}\label{alg:FISTA} \begin{enumerate}
\item Initialize $x^1 = 0$, $\omega = 0$, $\iterc = 0$, $s_1 = 1$, compute $ d^1 = \nabla g(x^1)$. Let $L$ be a Lipschitz constant for $g$.
\noindent \item While $\|\prox_{R}(\omega^\iterc  - d^\iterc )\| > \epsilon$
\begin{itemize}
\item Set $\iterc = \iterc+1$.
\item update $x^{\iterc} = \prox_{L^{-1} R}(\omega^{\iterc-1} - \alpha d^{\iterc-1})$.
\item set $s_\iterc = \frac{1 + \sqrt{1 + 4 s_{\iterc-1}^2}}{2}$
\item  set $\omega^\iterc  = x^\iterc  + \frac{s_{\iterc-1} - 1}{s_\iterc}(x_\iterc - x_{\iterc-1})$.
\item Compute $d^\iterc  = \nabla g(x^\iterc)$.
\end{itemize} \item Output $x^\iterc$.
\end{enumerate} \end{algorithm}

The iteration~\eqref{eq:prox_def} is known as the proximal gradient
iteration, and converges with the same rates as gradient descent on
the smooth function $g$ \citep{nesterov2013introductory}. The iteration
can be accelerated to achieve a better rate of convergence by using
an auxiliary iterative sequence; the most famous example of such an
algorithm is FISTA \citep{beck2009fast}. The FISTA algorithm is only
slightly more complicated than~\eqref{eq:prox_def}, and is detailed
in Algorithm~\ref{alg:FISTA}. 

Linear regression and logistic regression have easily computable Lipschitz
constants of $\|A\|^2$ and $\frac{1}{4}\|A^2\|$ respectively. Unfortunately,
the models for the Gamma and Poisson GLMs are constructed using logartithmic
and exponential $g_2$, and these functions do not have a global quadratic
upper bound. A safeguard linesearch is requried to ensure descent,
i.e., that the objective at $x^+$ is smaller than the objective at
$x^k$ unless $x^k$ is a stationary point. A FISTA with line search
replaces $L^{-1}$ in Algorithm$~\ref{alg:FISTA}$ with an iterative
step-size selection procedure to ensure

\[
g(x^{\iterc+1}) + R(x^{\iterc+1}) < g(x^{\iterc}) + R(x^{\iterc}). 
\]

However, in practice Algorithm~\ref{alg:FISTA} is a descent method
when $L$ is locally estimated at each iteration using a simple heuristic.
We discuss the heuristic and other specifics of the Gamma GLM in the
next section. 

\section{\label{sec:gamma} Adapting FISTA to the Gamma Family with EN Regularization}

In this section, we adapt the general scheme of fitting GLM model
with regularization term to the special case of GLM for gamma distributed
response variables with the elastic net regularization (glmGammaNet).
The optimization problem is as follows

\begin{equation}
\min_{x}NLL(x)+R_{EN}(x;\lambda,\,\alpha)\label{eq:glmGammaNet_Objective}
\end{equation}

where $NLL(x)$ is the negative log-likelihood, $\lambda$ is the
regularization parameter and $R_{EN}(x;\lambda,\,\alpha)$ is the
Elastic Net regularization term. The section proceeds as follows.
Section \ref{subsec: Gamma_NLL_ProxEN} derives the negative log-likelihood
(NLL), the gradient of the NLL and the proximal of the EN regularization
term. In Seciton \ref{subsec:Local-Approx-of-Lip-Const}, we develop
a customized method of obtaining a local approximation of the Lipschitz
constant to avoid a line search (which requires additional function
evaluations). Section \ref{subsec:CV_Lambda} shows how cross-validation
can be used to select the optimal regularization parameter and presents
the complete algorithm to fit GLM for gamma-distributed responses
with EN regularization.

\subsection{$NLL$, $\nabla NLL$ and Proximal of EN\label{subsec: Gamma_NLL_ProxEN}}

The probability density function of gamma distribution is given by

\[
f(b;k,\,\theta)=\dfrac{1}{\Gamma(k)\theta^{k}}b^{k-1}e^{-\dfrac{b}{\theta}},\;k>0,\;\theta>0,
\]

where $k$ is the shape parameter and $\theta$ is the scale parameter.
The expectation of a gamma random variable $B$ is given by

\[
\mathrm{E}(B)=k\theta.
\]

Using the logarithm link function, the relationship between the expectation
of $B_{i}$ and the linear compoment of GLM is given by

\[
\mathrm{E}(B_{i})=\mathrm{log}(k\theta_{i})=A_{i\cdot}x,
\]

where $k$ is the shape parameter, which is assumed to be the same
for all examples, $\theta_{i}$ is the scale parameter for the $i$th
example, $A_{i\cdot}$ is the $i$th row of the data matrix $A$,
$x$ is the coeffcients for the GLM model. Therefore, the scale parameter
can be written as

\[
\theta_{i}=e^{A_{i\cdot}x}/k.
\]

The objective function of problem \ref{eq:glmGammaNet_Objective}
over all examples is given by

\begin{align}
H(x;b,\,k,\,\lambda,\,\alpha)= & NLL(b;k,\,x)+R_{EN}(x;\lambda,\,\alpha)\nonumber \\
= & -\sum_{i=1}^{N}\text{log}f(b_{i};k,\,\theta_{i})+R_{EN}(x;\lambda,\,\alpha)\nonumber \\
= & \sum_{i=1}^{N}\log\Gamma(k)+k\log\theta_{i}-(k-1)\log b_{i}+\dfrac{b_{i}}{\theta_{i}}+R_{EN}(x;\lambda,\,\alpha)\nonumber \\
= & \sum_{i=1}^{N}\log\Gamma(k)+k\cdot A_{i\cdot}x-k\cdot\log k-(k-1)\log b_{i}+k\cdot b_{i}e^{-A_{i\cdot}x}+\lambda\left(\alpha\left\Vert x\right\Vert _{1}+\dfrac{1-\alpha}{2}\left\Vert x\right\Vert _{2}^{2}\right).\label{eq:Gamma_NLL}
\end{align}

The partial derivative of $NLL$ with respect to $x_{j}$ is

\begin{equation}
\dfrac{\partial NLL}{\partial x_{j}}=\sum_{i=1}^{N}k\cdot\left(1-b_{i}e^{-A_{i\cdot}x}\right)A_{ij}.\label{eq:Gamma_NLL_Gradient}
\end{equation}

In order to use FISTA, we also need the proximity operator (prox)
of the elastic regularization term

\[
R_{EN}(x;\lambda,\,\alpha)=\lambda\left(\alpha\left\Vert x\right\Vert _{1}+\dfrac{1-\alpha}{2}\left\Vert x\right\Vert _{2}^{2}\right)
\]

From \citep{parikh2014proximal}, the prox of $R_{EN}(x;\lambda,\,\alpha)$
is given by

\begin{equation}
\mathrm{prox}_{tR_{EN}}(v)=\dfrac{1}{1+t\lambda\left(1-\alpha\right)}\mathrm{sgn}\left(v\right)\mathrm{max}\left(\left|v\right|-t\lambda\alpha,\,0\right),\label{eq:EN_Prox}
\end{equation}

where $v$ is a vector, while $\mathrm{sgn(v)}$ and $\mathrm{max}(v)$
act on $v$ element-wise. 

\subsection{Computing a Local Upper Bound \label{subsec:Local-Approx-of-Lip-Const}}

Algorithm ~\ref{alg:FISTA} requires a global Lipschitz constant for
the gradient of the objective. However, the gradient \ref{eq:Gamma_NLL_Gradient}
does not have a global Lipschitz constant, because the exponential
function cannot have a global quadratic upper bound. A local quadratic
approximation can be computed efficiently and used in lieu of a global
Lipschitz constant. The local quadratic upper bound for the Gamma
model (\ref{eq:Gamma_NLL}) is given by 

\begin{equation}
L(x)=\|A\|_{F}^{2}\left(\sum_{i=1}^{N}k^{2}\cdot\left(1-b_{i}e^{-A_{i\cdot}x}\right)^{2}\right).\label{eq:LocalBound}
\end{equation}

The idea behind (\ref{eq:LocalBound}) is to get a data-dependent
local quadratic upper bound, analogous to those we have for linear
and logistic regression. The bound we use is conserviative in practice,
since we never need to activate the safe-guard line search. However,
it cannot be too conservative, as we see fast performance across the
testbed of problems, as discussed in the numerical expriments. 

\subsection{Optimal $\lambda$ via Cross Validation\label{subsec:CV_Lambda}}

With the results from section (\ref{subsec: Gamma_NLL_ProxEN}) and
(\ref{subsec:Local-Approx-of-Lip-Const}), we are able to solve the
optimzatio problem (\ref{eq:glmGammaNet_Objective}) if $\lambda$
and $\alpha$ are given. The choice of $\lambda$ and $\alpha$ has
a strong impact on the problem. Following the suggestions by \citep{friedman2010regularization},
we assume that $\alpha$ is determined by the user and focus on using
cross-validation to find the optimal value for $\lambda$. The cross-validation
procedure is as follows:
\begin{enumerate}
\item Compute the smallest $\lambda$ that gives an all-zero solution for
the regularized problem, call this $\lambda_{max}$.
\item The lower bound of the grid is given by $\lambda_{min}=\epsilon*\lambda_{max}$,
where $\epsilon$ is a user-defined constant with a default value
of 0.001.
\item The grid of $\lambda$ consists of $n_{\lambda}$ values between $\lambda_{min}$
and $\lambda_{max}$ equally spaced in the log scale, where $n_{\lambda}$
is a user-defined constant with a default value of 100.
\item For each $\lambda$, perform n-fold cross validation and compute the
mean NLL, where n is a user-defined constant with a default value
of 10. 
\item Choose the $\lambda$ with the smallest NLL for use in the final model.
\end{enumerate}

\subsubsection*{Computing $\lambda_{max}$}

The value $\lambda_{max}$ should be such that $x_{k}=0$ satisfies
the optimality conditions for the problem. Denoting the $j$th element
of $x_{k}$as $x_{k}^{j}$, optimality is equivalent to a fixed point
condition across all $j$: 

\begin{align*}
x_{k}^{j}= & \mathrm{prox}_{\frac{1}{L}R_{EN}}(x_{k}^{j}-\frac{1}{L}\nabla NLL^{j})\\
= & \dfrac{1}{1+\frac{1}{L}\lambda\left(1-\alpha\right)}\mathrm{sgn}\left(x_{k}^{j}-\frac{1}{L}\nabla NLL^{j}\right)\mathrm{max}\left(\left|x_{k}^{j}-\frac{1}{L}\nabla NLL^{j}\right|-\frac{1}{L}\lambda\alpha,\,0\right)
\end{align*}

The definition of $\lambda_{max}$ requires the fixed point condition
above to hold when $x_{k}^{j}=0$ for all $j$, so we have 

\begin{align}
0= & \mathrm{prox}_{\frac{1}{L}R_{EN}}(-\frac{1}{L}\nabla NLL^{j})\nonumber \\
= & \dfrac{1}{1+\frac{1}{L}\lambda\left(1-\alpha\right)}\mathrm{sgn}\left(-\frac{1}{L}\nabla NLL^{j}\right)\mathrm{max}\left(\left|\frac{1}{L}\nabla NLL^{j}\right|-\frac{1}{L}\lambda\alpha,\,0\right)\label{eq:prox_update_zero}
\end{align}

A sufficient condition to ensure above is that 

\[
\lambda\alpha\geq\left|\nabla NLL^{j}\right|=\left|\sum_{i=1}^{N}\left(k-\dfrac{b_{i}}{\theta_{i}}\right)A_{ij}\right|=\left|\sum_{i=1}^{N}\left(k-\dfrac{b_{i}}{1/k}\right)A_{ij}\right|=\left|\sum_{i=1}^{N}k\left(1-b_{i}\right)A_{ij}\right|
\]

To make sure that $x_{k}=0$ satisfies the optimality condition across
all $j$, we take $\lambda_{max}$ to be the largest of all such $\lambda$:

\[
\lambda_{max}=\max_{j}\left|\sum_{i=1}^{N}k\left(1-b_{i}\right)A_{ij}\right|.
\]

\subsection{GLM for Gamma Response Variables with Elastic Net (glmGammaNet)\label{subsec:GLM-for-Gamma_Alg}}

Algorithm \ref{alg:GLMRegModelSelection} gives the complete pseudo
code for glmGammaNet.

\begin{algorithm} \caption{glmGammaNet}\label{alg:GLMRegModelSelection} \begin{enumerate}
\item Set $A$, $b$, $\lambda_{max}$, $\epsilon$ and $n_{\lambda}$
\item Compute $\lambda_{max}=\max_{j}\left|\sum_{i=1}^{N}k\left(1-b_{i}\right)A_{ij}\right|$
\item Compute $\lambda_{min} = \epsilon*\lambda_{max}$
\item Compute vector of candidate $\lambda$'s, $\lambda_{vec} = \exp (\mathrm{seq}(\log(\lambda_{min}), \log(\lambda_{max}), \mathrm{length} = n_{\lambda}))$
\item For $j = 1:n_{\lambda}$
\begin{itemize}
\item $\lambda = \lambda_{vec}(j)$
\item for $i = 1:k$
\begin{itemize}
\item randomly divide $A$ into $k$ partitions. Let $A_{test}$ be one of the partitions, $A_{train}$ be the union of the rest of the partitions.
\item randomly divide $b$ into $k$ partitions. Let $b_{test}$ be one of the partitions, $b_{train}$ be the union of the rest of the partitions.
\item use the FISTA algorithm on the model defined $A_{train}$, $b_{train}$ and $\lambda$ to find the solution $x_{train}$
% \item Compute $b_{pred} = \exp(A_{test}^T x_{train})$
\item $\mathrm{NLL}_{ij} = NLL(x_{train}, A_{test}, b_{test}$
\end{itemize}
\item $\mathrm{NLL}_{j} = \sum_i \mathrm{NLL}_{ij}$
\end{itemize}
\item $\lambda_{best}$ is the $\lambda$ that results in the smallest value among all $\mathrm{NLL}_{j}$ \label{alg:LambdaChoice}
\item use the FISTA algorithm on the model defined $A$, $b$ and $\lambda_{best}$ to find the solution $x_{best}$
\end{enumerate} \end{algorithm}

\subsubsection*{Two alternative ways of choosing the best $\lambda$}

In Algorithm \ref{alg:GLMRegModelSelection}, we choose the $\lambda$
that minimizes the $NLL$, which seems to be the optimal choice. However,
this is not always the case. It is worth noting that the $NLL$ computed
in Algorithm \ref{alg:GLMRegModelSelection} are just estimates of
the true prediction errors. Therefore, there are uncertainies associated
with these estimates. To account for these uncertainties, we propose
two alternative ways of choosing the ``best'' $\lambda$. The first
alternative is to choose the maximum $\lambda$ with the corresponding
$NLL\leq NLL_{min}+SD_{NLL_{min}}$. The second alternative is to
choose the maximun $\lambda$ with the corresponding $NLL$ smaller
than the $\alpha$th percentile of all the $NLL$s from cross-validation.
We discuss the performance of both alternatives along with Algorithm
\ref{alg:GLMRegModelSelection} in the numerical experiment section.

\section{\label{sec:sim} Numerical Experiment}

We use Monte Carlo (MC) simulations to demonstrate the superior performance
of the three variants of glmGammaNet compared with the standard GLM
method. glmGamma is the standard GLM for gamma responses. glmGammaNet
is our method described in Algorithm \ref{alg:GLMRegModelSelection}.
glmGammaNet.percentile is the percentile variant described in section
\ref{subsec:GLM-for-Gamma_Alg} with 10th percentile threshold, and
glmGammaNet.percentile.nonzero is the result of fitting the glmGamma
without the zero coefficients identified by glmGammaNet.percentile.
glmGammaNet.1sd is the one-standard-deviation variant described in
section \ref{subsec:GLM-for-Gamma_Alg}, and glmGammaNet.1sd.nonzero
is the result of fitting the glmGamma without the zero coefficients
identified by glmGammaNet.1sd. We run 1000 MC simuations. In the $i$th
MC run, the following steps are performed:
\begin{enumerate}
\item Set $n=100$ and $p=15$, where $n$ is the number of examples and
$p$ is the dimension of the coeffcient vector $x$.
\item Generate the $n\times p$ predictor matrix $A$ from i.i.d. normal
distribution $N\sim(0,1)$.
\item Generate vector of length $p$ from i.i.d. normal distribution $N\sim(0,1)$
and randomly set 10 of the elements to zero. Denote the resulting
vector as $x_{true}$.
\item Compute vector of the true responses $b_{true}=\exp(Ax_{true})$.
\item Compute the vector of rates $\lambda_{true}=k/b_{true}$.
\item Generate $n\times1$ vector of response variables $b$ by getting
one sample from the gamma distribution specified by each element in
the vector $\lambda_{true}$. 
\item Use different methods to compute the solution $x_{METHOD}^{(i)}$
using $A$, $b$ as the input and save $x_{METHOD}^{(i)}$.
\end{enumerate}

\subsection{Error of Fitted Coeffcients}

We compute the following performance metrics for the error of the
fitted coeffcients
\begin{enumerate}
\item L1 Norm of the difference between $x_{METHOD}$ and $x_{true}$\\
\[
\mathrm{error.L1}_{METHOD}=\left\Vert x_{METHOD}-x_{true}\right\Vert 
\]
\item $\mathrm{error.L1}_{METHOD}$ as a percentage of $\left\Vert x_{true}\right\Vert $\\
\[
\%\,\mathrm{error.L1}_{METHOD}=\left\Vert x_{METHOD}-x_{true}\right\Vert /\left\Vert x_{true}\right\Vert \cdot100
\]
\end{enumerate}
Table \ref{tab:L1-Error} summarizes the error of fitted coefficients
for different GLM methods.

\begin{table}[th]
\begin{centering}
\begin{tabular}{|c||c||c|}
\hline 
 & error.L1 & \% error.L1\tabularnewline
\hline 
\hline 
glmGamma & 0.41 & 10.4\tabularnewline
\hline 
\hline 
glmGammaNet & 0.37 & 9.3\tabularnewline
\hline 
\hline 
glmGammaNet.percentile & 0.36 & 9.2\tabularnewline
\hline 
\hline 
glmGammaNet.1sd & 0.55 & 14.0\tabularnewline
\hline 
\hline 
glmGammaNet.percentile.nonzero & 0.33 & 8.3\tabularnewline
\hline 
\hline 
glmGammaNet.1sd.nonzero & 0.23 & 5.8\tabularnewline
\hline 
\end{tabular}
\par\end{centering}
\caption{Performance Summary of Different GLM methods\label{tab:L1-Error}}
\end{table}

As shown in Table \ref{tab:L1-Error}, the glmGamma method does a
reasonably good job, with a percentage error of 10.4\%. By adding
Elastic Net regularization and choosing the $\lambda$ with the smallest
NLL in cross-validation, we get a slight improvement in percentage
error, down to 9.3\%. The glmGammaNet.percentile method has a similar
percentage error of 9.2\%. The glmGammaNet.1sd method has the highest
percentage error of 14\%. However, if we drop the zero coefficients
identified by glmGammaNet.percentile and glmGammaNet.1sd, and then
perform a regular glmGamma, the percentage errors drop sifginicantly.
The glmGammaNet.percentile.nonzero method has a percentage error of
8.3\%, down from 9.2\% and the glmGammaNet.1sd.nonzero method has
a percentage error of 5.8\%, down from 14\%. We conjecture that the
reduction in percentage error is due to the variable selection power
of our new methods. In the next subsection we explore the variable
selection performance of different methods.

\subsection{Variable Selection Performance Analysis}

We compare the variable selection performance of different GLM methods
by examining the following two statistics:
\begin{enumerate}
\item Number of correctly identified zero coefficients, denoted as $\mathrm{zeros.correct}_{METHOD}$
\item Number of correctly identified nonzero coefficients denoted as $\mathrm{nonzeros.correct}_{METHOD}$
\end{enumerate}
Table \ref{tab:Variable-Selection-Results} summarizes the variable
selection performance for different GLM methods.

\begin{table}[H]
\begin{centering}
\begin{tabular}{|c||c||c|}
\hline 
 & zeros.correct & nonzeros.correct\tabularnewline
\hline 
\hline 
glmGamma & 0 & 5\tabularnewline
\hline 
\hline 
glmGammaNet & 1.976 & 5\tabularnewline
\hline 
\hline 
glmGammaNet.percentile & 4.771 & 5\tabularnewline
\hline 
\hline 
glmGammaNet.1sd & 7.815 & 5\tabularnewline
\hline 
\end{tabular}
\par\end{centering}
\caption{Performance Summary of Different GLM methods\label{tab:Variable-Selection-Results}}
\end{table}

All the methods studied have successfully included the nonzero coefficients
in their solution, yet the number of identified zero coefficients
varies a lot. Out of the 10 zero coefficients in the true solution,
the glmGamma method fails to identify any zero coefficients. On average,
the glmGammaNet method manages to find roughly 2 zero coefficients.
The glmGammaNet.percentile method identifies approximately 5 zero
coefficients, which presents a great improvement. Most notably, the
glmGammaNet.1sd method find an incredible 8 zero coefficients on average
and 80\% chance of correctly identifying 7 or more zero coefficients.
By dropping the variables with zero coefficients, we are essentially
removing noise in the dataset, therefore, the percentage error of
these methods are much better, as shown in the previous section.

\begin{figure}[H]
\begin{centering}
\includegraphics[width=0.8\textwidth]{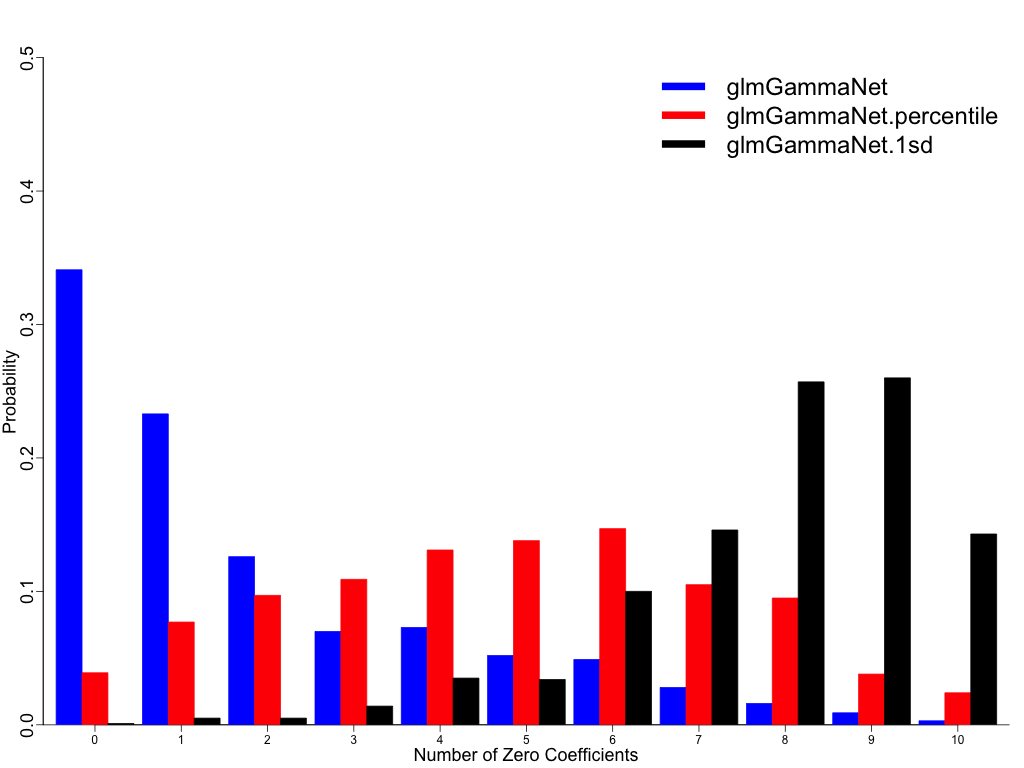}
\par\end{centering}
\caption{Histogram of number of zero coefficients selected over 1000 simulations\label{fig:Number-Nonzero-Histogram}}

\end{figure}

Figure \ref{fig:Number-Nonzero-Histogram} further visualizes the
distribution of the number of correctly identified zero coefficients.
The blue bars represent the results for glmGammaNet. Notice that there
is a 35\% chance that glmGammaNet does not identify any zero coefficients
and the probability decreases as the number of correctly identified
zero coefficients increases. This shows that glmGammaNet is very conservative
in terms of variable selection. The red bars show the results for
glmGammaNet.percentile. The distribution is roughly bell-shaped and
the peak occurs at 6 zero coefficients with a probability of 15\%.
This indicates that glmGammaNet.percentile is considerably more aggresive
than glmGammaNet, but still not satisfactory. The black bars show
the results for glmGammaNet.1sd. The distribution is concentrated
around 8 and 9 zero coefficients, which account for more than 0.5
probability. This shows that the glmGammaNet.1sd method performs extremely
well in variable selection.

\section{\label{sec:disc} Discussion}

In this paper, we developed a parallel implementation for GLM fitting
with Gamma distributed data and elastic net regularization. One reason
the Gamma may not be available in standard software is that the objective
function is a composition of an exponential model with a linear map,
and so does not have a global quadratic upper bound. We developed
a customized accelerated proximal gradient method by using local quadratic
estimates; although a safeguard line search is implemented, it was
never activated across our entire suite of experiments. We also provide
a straightforward cross-validation scheme to determine the optimal
value of the regularization parameter. Numerical experiments are show
the advantage of these methods over standard GLM without regularization.
The new methods have both smaller error in fitted coefficients and
superior variable selection performance. The choice of regularization
parameter is very important, and we recommend two simple strategies:
(1) conservative: using the parameter that corresponds to the smallest
negative loglikelihood in the cross validation, and (2) aggressive:
using the one-standard-deviation rule.

\nocite{}
\printbibliography
\end{document}